\documentclass[aps,prl,preprint,superscriptaddress,nofootinbib,]{revtex4}
\usepackage{setspace}
\pdfoutput=1
\usepackage{graphicx}
\begin{document}

\title{Structural Disjoining Potential for Grain Boundary Premelting and Grain Coalescence from Molecular-Dynamics Simulations}

\author{Saryu Fensin}
\affiliation{Department of Chemical Engineering and Materials Science, University of California, Davis, CA}
\author{David Olmsted}
\affiliation{Department of Physics, Northeastern University, Boston, MA}

\author{Dorel Buta}
\affiliation{Department of Chemical Engineering and Materials Science, University of California, Davis, CA}

\author{Mark Asta}
\affiliation{Department of Materials Science and Engineering, University of California, Berkeley, CA}
\affiliation{Department of Chemical Engineering and Materials Science, University of California, Davis, CA}
\author{Alain Karma}
\affiliation{Department of Physics, Northeastern University, Boston, MA}
\affiliation{Center for Interdisciplinary Research on Complex Systems, Northeastern University, Boston, MA}
\author{J. J. Hoyt}
\affiliation{Department of Materials Science and Engineering, McMaster University, Hamilton, ON, Canada}

\begin{abstract}
We describe a molecular dynamics framework for the direct calculation of the short-ranged 
structural forces underlying grain-boundary premelting and grain-coalescence in 
solidification.  The method is applied in a comparative study of (i) a 
$\Sigma$9 $\left\langle{115}\right\rangle$ 120$^o$ twist and (ii) a 
$\Sigma$9 $\left\langle{110}\right\rangle$ $\left\{411\right\}$ symmetric tilt boundary 
in a classical embedded-atom model of elemental Ni.  Although both boundaries feature 
highly disordered structures near the melting point, the nature of the temperature 
dependence of the width of the disordered regions in these boundaries is qualitatively 
different.  The former boundary displays behavior consistent with a logarithmically 
diverging premelted layer thickness as the melting temperature is approached from below, 
while the latter displays behavior featuring a finite grain-boundary width at the melting 
point. ÊIt is demonstrated that both types of behavior can be quantitatively described 
within a sharp-interface thermodynamic formalism involving a width-dependent interfacial 
free energy, referred to as the disjoining potential. The disjoining potential for 
boundary (i) is calculated to display a monotonic exponential dependence on width, while 
that of boundary (ii) features a weak attractive minimum.  The results of this work are 
discussed in relation to recent simulation and theoretical studies of the thermodynamic 
forces underlying grain-boundary premelting.

\end{abstract}
\vspace{1cm}

\maketitle

\section{Introduction}

At high homologous temperatures the atomic structure of a grain boundary often 
displays pronounced disorder.  In some cases this structural disorder can involve the 
formation of nanometer-scale intergranular films with liquid-like properties below the 
bulk melting point, a phenomenon commonly referred to as grain-boundary premelting.   
The interfacial thermodynamic driving forces underlying grain-boundary premelting are 
understood to be an important factor influencing grain coalescence behavior during the 
late stages of solidification (e.g., \cite{rappaz03}).  Specifically, when premelting 
is thermodynamically favored there exists an associated repulsive ``disjoining pressure" 
which hinders the coalescence of two misoriented grains at nanometer-scale distances.  
In such cases a significant undercooling may be required for dendrite arms to merge to 
form solid ``bridges" that are capable of sustaining thermal-contraction stresses 
without grain sliding or rupture.  The quantitative characterization of 
grain-boundary disjoining forces is thus an important issue in the context of modeling 
the formation of solidification defects, known as ``hot tears," which occur deep 
within the mushy zone during casting or welding \cite{rappaz99,rappaz03,wang04,asta09}.

Despite its practical importance in the context of solidification defects, the 
magnitude and spatial extent of grain-boundary 
disjoining forces in metals and alloys 
remain incompletely understood.  As in the case of surface premelting \cite{dash95}, 
these forces can in principle be probed experimentally through measurements of the 
extent of equilibrium premelting as a function of temperature near the bulk melting 
point.  In comparison to surface premelting, however, the challenges inherent in 
characterizing the structure of ``buried" internal interfaces at high homologous
temperatures have significantly limited the number of direct experimental studies of 
grain-boundary premelting \cite{baluffi, masamura, watanabe, vold, inoko, divinski, lou, gupta}.  This situation has provided motivation 
for several recent theoretical studies based on conventional phase-field 
\cite{lobkovsky02,tang06,mishin09,wang09} and phase-field-crystal (PFC) 
\cite{berry08,mellenthin08} methods, which have led to new insights into the 
rich variety of possible premelting behavior that may be exhibited by grain boundaries 
as a function of their bi-crystallography.

Due to the difficulties inherent in validating theoretical models for grain-boundary 
premelting directly from experimental measurements, the authors recently proposed an 
independent methodology for calculating grain-boundary disjoining forces from 
equilibrium molecular-dynamics (MD) simulations \cite{hoyt09}.  The technique 
represents an extension of the numerous previous MD studies of grain-boundary 
premelting performed over the past three decades \cite{nguyen86, nguyen85, ciccotti83, ciccotti84, brough86, brough98, lutsko89, kikuchi, suzuki, besold, wolf, lu, williams, kaski}.  The technique for 
calculating disjoining potentials by MD is based on an analysis of the equilibrium 
distribution of the widths of premelted intergranular films, which are related to
the underlying disjoining forces through the fluctuation-dissipation theorem.  
The purpose of this paper is to describe the technical details surrounding the 
implementation of this method, and to demonstrate its application in the study of two 
distinct classes of grain boundary premelting behavior.  Specifically, we consider 
the premelting behavior of a high and intermediate energy boundary in elemental Ni
where the widths of the premelted layers continuously increase or remain finite, 
respectively, as the melting point is approached from below.  It is shown that the 
former behavior is consistent with an interfacial free energy ($\Psi$) that decreases 
exponentially with increasing width ($w$) of the premelted layer, while the latter 
behavior can be quantitatively modeled with a dependence of $\Psi$ on $w$ that 
features a weak attractive minimum at nanometer-scale widths.  This minimum
is accompanied by a grain boundary structure with alternating solid bridges and
disordered regions that bears strong similarities to the type of structure found
to be associated with such a minimum in a recent phase-field-crystal study 
\cite{mellenthin08}.

The remainder of this paper is outlined as follows.  In the next section we review 
a continuum, sharp-interface formalism for the thermodynamic properties underlying
grain-boundary premelting and coalescence, based on the definition of the so-called 
``disjoining potential," the (negative) derivative of which is the disjoining 
pressure referred to above.  Section III discusses the technical details underlying 
the calculation of the disjoining potential by MD, as well as the details of the 
simulations undertaken in the current application of the method to the study of 
high-temperature grain boundaries in fcc Ni.  The results are presented in section 
IV, followed by a summary and discussion of the findings in the context of previous 
theoretical studies in section V.

\section{Disjoining Potential}

The equilibrium width of a premelted grain-boundary is governed by a competition 
between bulk and interfacial thermodynamic factors, which can be represented 
mathematically as follows:
\begin{equation}
G(w)=\Delta G_f w + \Psi(w).
\label{eq}
\end{equation}
In Eq.~\ref{eq}, $G(w)$ represents the total excess free energy of a premelted 
grain boundary of width $w$, which is composed of two contributions:  $\Delta G_f$ 
represents the bulk free energy difference between liquid and solid phases
(positive below the melting point), per 
unit volume, while $\Psi(w)$ represents the width-dependent interfacial free energy, 
which we refer to as the disjoining potential.  The function $\Psi(w)$ takes the 
limits of $\gamma_{GB}$ (the interfacial free energy of a hypothetical ``dry" grain 
boundary) and $2 \gamma_{SL}$ (twice the solid-liquid interfacial free energy) for 
small and infinite values of $w$, respectively.  For intermediate values of the 
width $\Psi(w)$ can display a complex dependence on $w$.

The dependence of $\Psi(w)$ on $w$ is generally governed by distinct short and 
long-ranged contributions.  An attractive interaction between solid-liquid 
interfaces \cite{lipowsky,Clarke} arises due to dispersion forces which are dominant 
at large $w$, and are predicted to give rise to finite interfacial widths at $T_M$ 
\cite{lipowsky}.  For systems where the wetting condition, 
$\gamma_{GB} >2 \gamma_{SL}$ holds, a repulsive contribution to $\Psi(w)$ arises 
from short-ranged structural interactions  ($\Psi_{sr}$) associated with the overlap 
of the density waves in the diffuse regions of the solid-liquid interfaces.  
Mean-field arguments \cite{Widom}, as well as lattice-gas models (e.g., 
\cite{kikuchi}), yield an exponentially decaying form for this short-ranged 
contribution:
\begin{equation}
\Psi_{sr}(w)=2 \gamma_{SL} + \Delta \gamma \exp [-w/\delta]
\label{eq4}
\end{equation}
where $\Delta \gamma = \gamma_{GB}-2\gamma_{SL}$ and $\delta$ is an interaction length 
on the order of the atomic spacing.  In the absence of long-ranged dispersion forces,
insertion of Eq.~\ref{eq4} into Eq.~\ref{eq} leads to the 
prediction of an equilibrium grain boundary width that diverges logarithmically as 
the melting temperature ($T_M$) is approached from below. In previous work
\cite{hoyt09} we have argued, based on previous estimates of the dispersion forces for 
surface premelting, that the structural contributions to the disjoining potential for 
grain-boundary premelting in metals are expected to dominate the long-ranged dispersion 
contributions for nanometer-scale grain-boundary widths.

Recent theoretical results demonstrate that $\Psi_{sr}(w)$ can generally display more 
complex dependencies on $w$ than suggested by Eq.~\ref{eq4}.  Diffuse-interface phase 
field models \cite{lobkovsky02,tang06,mishin09}, which neglect dispersion forces and thus 
model $\Psi_{sr}$ directly, have shown that depending on grain-boundary misorientation 
and the detailed choice of interfacial thermodynamic parameters, the premelting 
transition can exhibit either the continuous behavior associated with Eq.~\ref{eq4}, 
a hysteretic first-order character, or an intermediate behavior where the boundary
width increases with increasing temperature but remains finite at $T_M$ (these three
types of behavior are referred to as types 2, 1 and 3 in Ref.~\cite{kaski}).  The PFC 
method was recently used to study grain boundary premelting for both three-dimensional
body-centered-cubic (bcc) \cite{berry08} and two-dimensional hexagonal \cite{mellenthin08} 
systems.  The latter study involved a systematic investigation of symmetric tilt 
boundaries as a function of misorientation. Well beyond a critical misorientation angle 
(where $\gamma_{GB} > 2 \gamma_{SL}$) $\Psi_{sr}$ was found to exhibit a purely 
``repulsive" behavior, monotonically decreasing with increasing $w$, consistent with 
Eq.~\ref{eq4}.  However, well below the critical angle, $\Psi_{sr}$ exhibited an 
attractive minimum, giving rise to a finite area-averaged liquid-layer width at the 
melting temperature, reflecting the presence of localized premelting within the cores 
of the grain-boundary dislocations in low-angle boundaries.

The possibility that the short-ranged contributions to the disjoining potential can 
exhibit an attractive minimum was considered in earlier theoretical studies of wetting 
transitions \cite{lipkowsky87}.  In these studies a double-exponential ansatz was used 
to model such behavior.  For the purposes of the present study this double-exponential 
form can be written as follows:
\begin{equation}
\Psi_{sr}(w) = 2\gamma_{SL} + [\Delta_1 \exp{(-w/\delta_1)} - \Delta_2 \exp{(-w/\delta_2)}],
\label{eq5}
\end{equation}
where $\Delta_1$ and $\Delta_2$, which are both positive quantities, represent the 
strengths of repulsive and attractive exponential contributions with decay lengths 
$\delta_1$ and $\delta_2$, respectively.  Although Eq.~\ref{eq5} has not been derived
directly from a microscopic theory, it provides a useful phenomenological form to 
model the MD data, as shown below.  This equation gives rise to a predicted 
value of the quantity $\gamma_{GB} - 2\gamma_{SL}$ equal to $\Delta_1 - \Delta_2$, 
where again $\gamma_{GB}$ is the interfacial free energy of a hypothetical dry grain 
boundary given by the limit of $w$ going to zero in Eq.~\ref{eq5}.  Choosing 
$\delta_2 > \delta_1$, and if $\Delta_1> \Delta_2$, as in the results given below, 
this disjoining potential is repulsive at short distances, attractive at large $w$, 
and features a minimum at a width $w_m$ given as:
\begin{equation}
\label{eq_wm} 
w_m = \frac{\delta_1 \delta_2}{\delta_2 - \delta_1} \ln{[\frac{\Delta_1 \delta_2}{\Delta_2 \delta_1}]}. 
\end{equation}
With the form for the disjoining potential given by Eq.~\ref{eq5} the width of the 
premelted boundary is predicted to be finite, with a value $w_m$, at the melting 
point.

In the MD results presented below we demonstrate the disordering behavior for two grain boundaries with relatively 
high and intermediate energy  which we show can be quantitatively 
modeled by Eqs.~\ref{eq4} and \ref{eq5}, respectively.  To compute quantitative values 
for the disjoining potentials from these simulations, we make use of the following 
relation between $G(w)$ and the equilibrium distribution $P(w)$ of grain-boundary 
widths sampled by the MD systems at a temperature $T$: 
$P(w) \propto \exp{[-A G(w) / k_B T]}$, where $A$ is the interfacial area.  
Calculations of $P(w)$ by MD, combined with an accurate knowledge of the bulk melting 
properties underlying $\Delta G_f$ allows one to extract $\Psi(w)$, as described below.  
Since the MD simulations are based on a short-ranged classical interatomic-potential 
model, they do not include the long-ranged dispersion-force contributions to 
$\Psi(w)$, and thus sample $\Psi_{sr}(w)$ directly.

\section{\label{Methods}Methodology}

In this study the simulations were based on an embedded-atom method (EAM)
interatomic potential model for Ni developed by Foiles, Baskes and Daw (FBD)
\cite{FBD}.  This potential was chosen as we have in previous work characterized 
both the bulk melting properties and solid-liquid interfacial free energies for FBD Ni; 
both sets of properties are prerequisites for a quantitative study of grain-boundary
premelting.  As reviewed in Ref.~\cite{hoyt09} the melting temperature for the FBD
Ni potential has been bracketed to lie in the range T$_M$=1709-1710 K, and the 
value of the solid-liquid interfacial free energy has been calculated to be 
$\gamma_{SL}$=285 mJ/m$^2$ \cite{hoyt03}. The latent heat for the potential was also calculated to be 0.015  $eV/$\AA$^3$.

In the present study we considered the following three different grain boundary
structures: a
$\Sigma$9 symmetric $\left\langle{0 1 1}\right\rangle$$\left\{4 1 1\right\}$ 38.9$^o$ 
tilt boundary (hereafter referred to as the $\Sigma$9 tilt boundary), a
$\Sigma$11 symmetric $\left\langle{0 1 1}\right\rangle$$\left\{3 1 1\right\}$ 50.5$^o$ 
tilt boundary (hereafter referred to as the $\Sigma$11 tilt boundary), 
and a $\Sigma$9 $\left\langle{1 1 5}\right\rangle$ 120$^o$ twist boundary (hereafter 
referred to for convenience as the $\Sigma$9 twist boundary, even though we recognize 
that this boundary can also be described as a symmetric tilt boundary).  These were 
selected from a large library of grain boundary structures that fit into moderately-sized 
periodic simulation cells, developed in the work of Olmsted  et al.\cite{Olmsted}.  As 
shown in Table~\ref{tab1}, the selected boundaries have zero-temperature energies 
($\gamma^0_{GB}$) that span a range of values relative to 2$\gamma_{SL}$, and were thus 
expected to feature a range of premelting behavior.

\renewcommand{\baselinestretch}{1.5}
\begin{table*}[top]
\centering
\begin{tabular}{  l c |cc |c| c| c| c}
\hline
\hline
Grain Boundary type && & $L_{x}$ (\AA) & $L_{y}$ (\AA) & $L_{z}$ (\AA) & $\gamma^0_{GB}$ ($mJ/m^2$)& $\gamma^0_{GB}/2\gamma_{SL}$ \\
 \hline
 \hline
$\Sigma$9 $\left\langle{0 1 1}\right\rangle$ $\left\{4 1 1\right\}$ 38.9$^o$ symmetric tilt   &&&  237.86    &   32.36 & 31.68 &   909    & 1.5  \\

 $\Sigma$9 $\left\langle{1 1 5}\right\rangle$ 120$^o$ twist  &&& 253.23 & 37.34    &   38.80  & 1440 & 2.5\\

$\Sigma$11  $\left\langle{0 1 1}\right\rangle$$\left\{3 1 1\right\}$ 50.5$^o$ symmetric tilt   &&&  246.55  & 32.36 &  33.02 & 450 & 0.79 \\
\hline
\hline
\end{tabular}
\renewcommand{\baselinestretch}{1.0}
\caption[prop table]
{The grain boundary types used in the MD simulations along with the simulation cell sizes, 
reported at zero temperature.  For reference the lattice constant of FBD Ni potential at 
zero temperature is 3.52 $\AA$.  These boundaries span a large range of 
$\gamma^0_{GB} / 2 \gamma_{SL}$, where $\gamma^0_{GB}$ is the zero-temperature grain-boundary
energy, and $\gamma_{SL}$ is the solid-liquid interfacial free energy.}
\label{tab1}
\end{table*}

\subsection{Energy Minimization}
\begin{figure}[top]
\includegraphics[scale=.8]{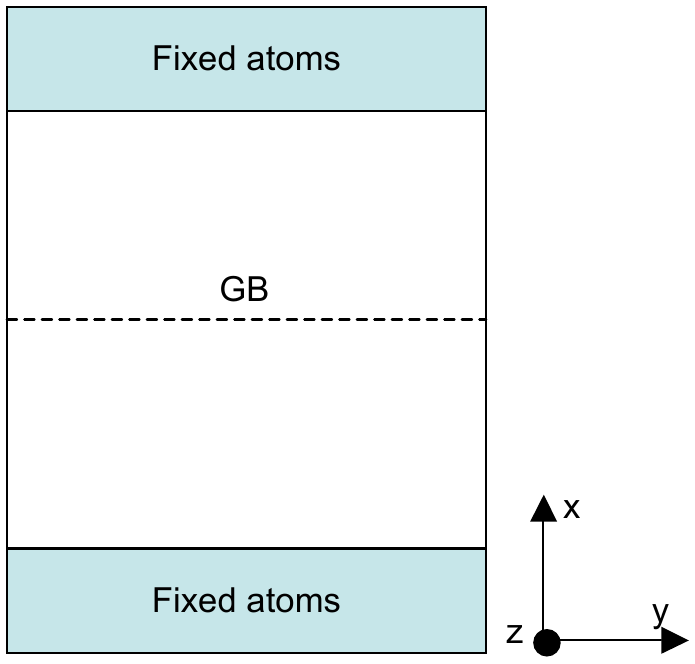}
\caption{\label{cell} Computational cell used for molecular dynamics simulations at zero temperature. The boundary plane is in the y-z directions. The cell is periodic in y and z.}
\end{figure}

The optimization of grain-boundary structures at zero temperature made use of a simulation block with two grains separated by a flat grain boundary as shown in Fig~\ref{cell}. 
The cell was periodic in the plane of the boundary and non-periodic perpendicular to it. 
The grains were sandwiched between two slabs parallel to the grain-boundary plane. The atoms in 
each of these slabs were fixed relative to each other and could only undergo a rigid body motion. The slabs could move normal to the boundary, allowing volume expansion to maintain
zero normal stress, and in the plane of the boundary, avoiding any resistance from the 
surfaces to translational movement of the grains relative to each other.  Multiple trial 
configurations were built in order to minimize the boundary energy with respect to 
relative translations of the grains and with respect to the number of atoms in the grain boundary, as 
described in Ref.~\cite{Olmsted}.   The energy of each trial configuration was minimized 
using the conjugate gradient method.  After minimization, the energy of  the grain 
boundary was computed as the total energy of the unconstrained atoms, less the bulk 
crystal energy for the same number of atoms, divided by the area of the boundary.  The 
dimensions of the computational cell used in these energy minimizations and in the 
subsequent MD simulations are given in Table~\ref{tab1}.

\subsection{Molecular-Dynamics Simulations}
Finite temperature simulations were performed by MD employing the LAMMPS code 
\cite{plimpton}.   All simulations were performed with a constant number of atoms, constant volume and temperature (NVT ensemble).  The temperature was maintained by using a Nose-Hoover thermostat \cite{nost} with a thermostat relaxation time of 0.1 ps, employing a time-step of 1 fs.  The simulations began with the optimized zero-temperature geometry, which was 
equilibrated at different  temperatures.  Prior to each simulation at a given 
temperature, the periodic lengths parallel to the grain boundary plane were 
expanded according to the finite-temperature value of the lattice constant for the bulk 
crystal (determined separately). These periodic lengths were then held constant in all 
of the NVT MD simulations.  The atoms in the slabs that were held fixed during the 
energy minimization procedure were made dynamic for the finite-temperature 
simulations, and the boundary conditions were modified such that both the grains 
terminated in free surfaces.

As discussed in the next sub-section, two types of analysis were conducted on the 
simulated systems to study their premelting behavior.  The first was a calculation of the 
excess volume as a function of temperature.  In the simulations used for this analysis equilibration times were a few ns.  The system volume was sampled over a total time of at least 10 ns at a frequency of 10 ps.

The second analysis involved the calculation of equilibrium grain boundary width histograms.  For 
these analyses, the simulations started at a given temperature with an equilibration 
lasting 10 ns.  Statistics were then obtained for the boundary width histograms from 
a total of 4000 snaphsots, sampled at a frequency of 10 ps.  As discussed below,
this number of snapshots and sampling rate ensured that at least a hundred statistically
independent samples were obtained at each of the temperatures studied for the 
histogram analysis.

\subsection{\label{EV}Calculation of Excess Volume and Width Histograms}
Prior to performing detailed analyses of width histograms, we performed an analysis 
of premelting behavior based on the temperature dependence of the excess volume.  
The excess volume was calculated by time averaging over values for single 
configurations, computed as follows.  For a single snapshot the excess volume was 
computed from the distance between two lattice planes, one in each grain.  These 
planes were chosen in each grain such that they were far from the boundary and the 
free surfaces.  The volume of material that would lie between the two planes in a
perfect crystal can be computed by counting the total number of atoms between the 
two planes (including half of the atoms in each of the two planes) and multiplying 
that number by the volume per atom of the bulk crystal at the same temperature (and 
zero pressure).  The difference between the actual volume and the expected bulk 
volume, divided by the area of the boundary, is the excess volume.  A slight linear 
dependence of this measured excess volume on the distance between the planes
was found, presumably the result of the numerical error in the lattice constant at high 
temperature.  However, any consistent choice was adequate for our purposes here, 
as the excess volume results were used mainly to determine the qualitative nature 
of the premelting behavior, as discussed below.  The specific planes chosen for the
excess volume were 1/4 and 3/4 of the way through the simulation cell.

In order to compute equilibrium grain-boundary width distributions, $P(w,T_i)$, for a given 
interface temperature $T_i$, we proceeded as follows.  For each snapshot the 
grain-boundary width $w$ was determined using a scheme developed by Hoyt et 
al. \cite{hoyt03} for the analysis of solid-liquid interface capillary fluctuations.  In 
this approach each atom is assigned a structural order parameter, 
$\phi_i=\frac{1}{12}\sum_{j}{\mid\vec{r}_{ij}-\vec{r}^{c}_{ij}\mid}^2$, where $r_{ij}$ 
are the actual positions of the 12 nearest neighbors of atom $i$ and $r^{c}_{ij}$ are 
the atom sites for the corresponding neighbors in the perfect crystal. The $\phi_i$ 
values are then averaged in bins along the direction normal to the boundary and 
the point of inflection in the average order parameter profile is taken as the 
position of the grain boundary. In the case of grain boundaries two separate 
profiles are required.  The first uses $r^{c}_{ij}$ for the crystal orientation of one of
the two grains in the bicrystal, and for the second $r^{c}_{ij}$ is chosen based on 
the other grain.  After these two order parameter profiles ($\phi_1$  and $\phi_2$) 
are obtained each is fit to the following function:
\begin{equation}
\phi(x)=a+b\tanh{[(x-x_0)/d]},
\label{fit}
\end{equation}
where \textit{a}, \textit{b}, \textit{$x_0$} and \textit{d} are fitting parameters, and the boundary width is
derived from the difference in values of $x_0$ determined from the $\phi_1$ and 
$\phi_2$ profiles.

\begin{figure}[top]
\includegraphics[scale=.5]{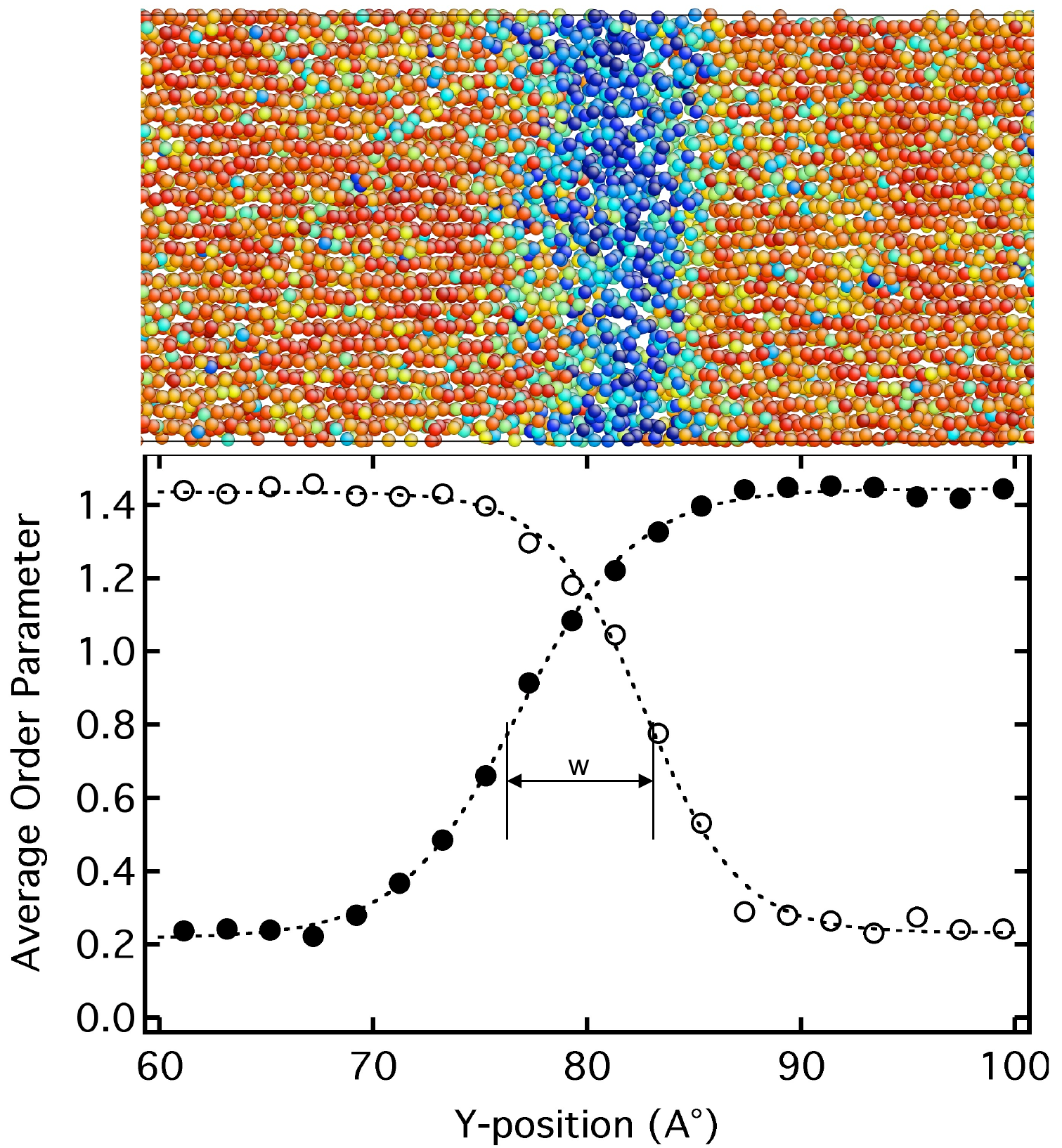}
\caption{\label{prof} A snapshot of a premelted grain boundary at an undercooling of 2$K$ and the 
corresponding average order parameter versus distance across the grain boundary.  Atoms in the 
top panel are colored based on the order parameter value with lighter color (red in the on-line figure) representing 
atoms in either of the two grains and darker color (blue in the on-line figure) showing atoms in the grain boundary. The construction to determine 
the grain-boundary width $w$ is illustrated in the bottom panel and a width of $w \approx 13.5 \AA$ 
is obtained.}
\end{figure}

The analysis used to compute grain-boundary widths is illustrated graphically in 
Fig.~\ref{prof} for a snapshot of  the $\Sigma$9 twist boundary at an undercooling 2 K below the bulk melting temperature.  In 
Fig.~\ref{prof} the top panel shows the atoms, color coded according to the 
magnitude of the difference in the order parameter values, lighter colors (red in the on-line figure) denoting atoms in 
an environment corresponding to either of the two grains, and darker colors (blue in the on-line figure) indicating an 
environment that does not correspond to either grain.  The corresponding order 
parameter profiles with the fits to Eq.~\ref{fit} (solid lines) are shown in the bottom 
of Fig.~\ref{prof}. The width is defined as the difference in the center positions of 
the fits to $\phi_1$ and $\phi_2$. In this example, a width of 13.5 $\AA$ is 
obtained by the analysis.

It should be emphasized that the choice of an order parameter used to compute the grain
boundary width is not unique (this point was discussed also in a recent related study by 
Williams and Mishin \cite{williams}).  Other choices for the structural order 
parameter used to define an interface width would include the so-called centrosymmetry 
parameter \cite{centro}, the order-parameter used by Morris in studies of solid-liquid 
interfaces \cite{morris02}, excess mass used by Mellenthin\cite{mellenthin08}, a parameter introduced by 
Williams and Mishin \cite{williams} based on the local structure factor, and an
order parameter introduced by von Alfthan et al. \cite{kaski} based on structural units.  
Different choices for the order 
parameter are expected to lead to slightly different values for the interface width, and 
the quantitative values of the disjoining potential derived from them.   For example,
in the sharp-interface formalism we describe the limit of $\Psi(w)$ for small $w$ as the 
interfacial free energy of a hypothetical ``dry" grain boundary - we would expect that 
this limiting value will depend on the way in which the width is defined, which is not 
unreasonable since with any choice of this definition a real boundary at zero temperature 
will have some small finite width.  The ambiguities associated
with the different choices for the definition of width is inherent in the use of a 
sharp-interface theoretical formalism to describe the properties of systems such as these 
with diffuse interfaces. In applications of such sharp-interface models, however, the 
ambiguity presents no problem in practice provided that each of the bulk and interfacial 
thermodyamic quantities are defined consistently.

In order to determine whether the simulation times employed in this study were 
sufficient to give adequate statistics for the width histograms, we estimated the
correlation time for width fluctuations based on an analysis of the decay of a
width autocorrelation function $C(\Delta{t})$ defined as follows:
\begin{equation}
C(\Delta{t})=\left\langle{w(t)w(t+\Delta{t})}\right\rangle-\left\langle{w}\right\rangle^2,
\label{cr}
\end{equation}
In Eq.~\ref{cr},  $w(t)$ denotes the instantaneous width of the grain boundary at a time 
$t$, and $\left\langle{...}\right\rangle$ denote ensemble (time) averages.  The time 
required for the autocorrelation function to decay by a fraction of $1/e$ was taken as 
a measure of the correlation time.  

The correlation times were found to vary significantly with boundary type and 
temperature.  The maximum value of the correlation time was obtained as 100 ps for the 
$\Sigma$9 tilt boundary at a temperature 2 K below $T_M$.  Given that 4000 
snapshots were sampled in the MD simulations, at a frequency of 10 ps, at least a 
hundred independent samples were obtained for each of the histograms presented
below.

\subsection{\label{HA}Calculation of Disjoining Potentials from Width Histograms}
As discussed in the previous section, the probability of observing a given boundary 
width at an interface temperature T$_i$ is related to the total free energy of the system 
as follows:
\begin{equation}
P(w,T_{i})=C_{i} \exp [-AG(w,T_{i})/k_B T_{i}],
\label{eq1}
\end{equation}
where $G(w,T_{i})$ is given by Eq.~\ref{eq}. In the above expression the subscript $i$ 
denotes one of the seven different temperatures for which the width histograms have been 
calculated by MD:  $T_{1}=1710$, $T_{2}=1708$, $T_{3}=1705$, $T_{4}=1700$,
$T_{5}=1690$, $T_{6}=1680$, $T_{7}=1650 K$, which represent undercoolings relative
to the bulk melting temperature, $T_M$=1710 K, ranging between zero and 60 K.  
Equation~\ref{eq1} emphasizes the fact that each undercooling yields a histogram that 
spans a different range of widths, and the unknown normalization constants at each 
temperature are denoted as $C_i$.

As discussed in the next section, analyses of the calculated width histograms were 
undertaken to compute disjoining potentials for two different boundaries, one of which
(the $\Sigma$9 twist) featured a diverging interface width as $T_M$ is approach from 
below, and another (the $\Sigma$9 tilt) whose width remains finite at $T_M$.   In the former case the disjoining potential was modeled using the form 
given by Eq.~\ref{eq4}, while for the latter use was made of Eq.~\ref{eq5}.
The unknown parameters in these expressions 
can be obtained in two ways based on the MD-calculated width histograms using 
Eq.~\ref{eq1}.  The first is to fit the histogram data at each temperature independently.  
This gives a range of values for the disjoining potential parameters, which are expected 
to be more accurate at the higher temperatures where one has access to the largest 
range of width values.  A refined estimate of the potential parameters can be obtained 
from a histogram analysis using all of the data at once, as follows.

From Eq.~\ref{eq} the disjoining potential can be written as
\begin{equation}
\Psi(w) = G(w,T_i)-w \Delta G_f (T_i)
\label{eq2}
\end{equation}
where $\Delta G_f$ is the bulk free energy difference between liquid and solid per unit
volume. For the EAM Ni system studied here the temperature dependence of 
$\Delta G_f$ is known accurately from previous studies.  For the small range of 
temperatures considered in the MD simulations of width histograms, which are close 
to the bulk melting temperature, $\Delta G_f$ can be accurately computed from the 
following expression:
\begin{equation}
\Delta G_f = L (\Delta T_{i} / T_M) 
\label{eq3}
\end{equation}
where $L$ is the latent heat per unit volume, and $ \Delta T_i = T_M - T_i$ is the 
interface undercooling.  From Eq.~\ref{eq1}, the disjoining potential can be written in 
terms of the equilibrium width distribution as:
\begin{equation}
\Psi(w) = -(k_B T_i) ln P(w,T_i)/A_i - \Delta G_f w + a_i
\label{eq6}
\end{equation}
where the $a_i$ are unknown constants related to $C_i$ in Eq.~\ref{eq1}. These 
parameters lead to constant offsets when the quantity 
-$k_B T_i \ln{P(w,T_i)}/A_i - \Delta G_f w$ is plotted versus $w$ for each interface 
temperature.  To construct the disjoining potential a least squares fit is used to refine 
the values of the shift parameters ($a_i$) along with the potential parameters 
($\Delta \gamma$ and $\delta$ in Eq.~\ref{eq4} or $\Delta_1$, $\delta_1$, 
$\Delta_2 $, and $\delta_2$ in Eq.~\ref{eq5}).  In this procedure the initial values for 
the shift parameters were estimated ``by eye" to give maximal overlap between 
the data, and the initial values for the potential parameters were obtained from an average 
of the independent fits to the data at separate temperatures (described above).  The
values of the parameters were then iteratively refined to minimize the square of the
differences between the MD data and the analytical expressions for $\Psi(w)$.

\subsection{\label{details}Effects of Thermostat, System Size and Boundary Conditions}
In the course of this work additional studies 
were also conducted to investigate the effects of the details of the simulation 
methodology on the resulting disjoining potentials.  These additional analyses were 
performed for the $\Sigma$9 twist boundary in Ni as well as a series of
 $\left\langle{100}\right\rangle$ tilt boundaries in Ni and bcc Fe.

The first issue addressed in these additional simulations was the effect of thermostat.  
Equation~\ref{eq1} assumes that the measured grain-boundary widths sample a canonical 
ensemble, and in order to achieve this by MD simulations we employed a standard 
Nose-Hoover thermostat \cite{nost} in the present work.  An alternative choice would 
be to use a Langevin thermostat \cite{langevin}, which would relax temperature fluctuations faster 
locally and produce different dynamics in the system, but would be intended to sample 
the same ensemble.  To test for unexpected effects of dynamics on the equilibrium 
width histograms and disjoining potential, two boundaries showing continuous 
premelting behavior (i.e., diverging widths as the melting temperature is approached 
from below), one in bcc Fe and one in fcc Ni, were simulated using both a Nose-Hoover 
thermostat and a Langevin thermostat with a time constant of 0.25 ps.  No statistically significant differences in the disjoining potential 
were found with the two thermostats.  Further, we tested a range of reasonable 
time constants for both thermostats, finding that statistically significant effects on the
calculated disjoining potentials were again absent unless the relaxation time for the 
Langevin thermostat was reduced to extremely short time scales, on the order of the 
inverse of the vibrational frequencies in the system.  Interestingly, in simulations which 
used the same time constant of 0.25 ps, the correlation time for width fluctuations was 
found to be substantially reduced in the simulations with the Langevin relative to those
with the Nose-Hoover thermostat.  The Langevin thermostat thus offers the potential advantage 
of providing improved statistics related to width histograms for a given simulation 
time.

The second issue investigated was the effect of system size on the calculated 
disjoining potential.  For atomically rough grain boundaries, such as those investigated 
in this work, previous theoretical studies have demonstrated that capillary fluctuations 
should give rise to a weak dependence of the disjoining potential on cross-sectional 
area \cite{kikuchi, lobkovsky02}.  To check for any unexpected large system-size effects,
simulations were performed for the $\Sigma$9 twist boundary using both the 
cross-sectional area given in Table~\ref{tab1}, as well as a value that was four times 
larger (i.e, with periodic lengths that were doubled in each of directions parallel to the 
boundary).  Although the width histograms obtained for the larger system were much 
narrower, as expected from Eq.~\ref{eq1}, it was still possible to obtain data over a 
sufficiently large range of widths to extract a disjoining potential.  The results for the 
disjoining potential were found to be consistent with those obtained from the smaller 
system provided that a small effect of system size on the bulk melting temperature 
(amounting to a roughly one-degree lowering of  $T_M$ for the larger system, which is
within the accuracy of the known value of the bulk melting point) was accounted 
for in the analysis.  Similar results were obtained in analyses of system size effects on 
calculated disjoining potential for a high-energy $\left\langle{100}\right\rangle$ tilt 
boundary in Ni.

A final issue that was investigated in these studies concerns the effect of boundary 
conditions on the calculated width distributions.  For a high-angle 
$\left\langle{100}\right\rangle$ symmetric tilt 
boundary in Fe this was investigated by computing width distributions at 10 K undercooling 
using both the free-surface boundary conditions employed in the current study, as well as 
a full periodic boundary condition in a simulation cell containing two grain boundaries 
and a periodic length normal to the boundaries that was allowed to be dynamic to maintain 
zero normal stress.  No statistically significant differences were found between the width 
distributions derived with the two different boundary conditions.

\section{\label{Res}Results}

\begin{figure}[here]
\includegraphics[scale=0.23]{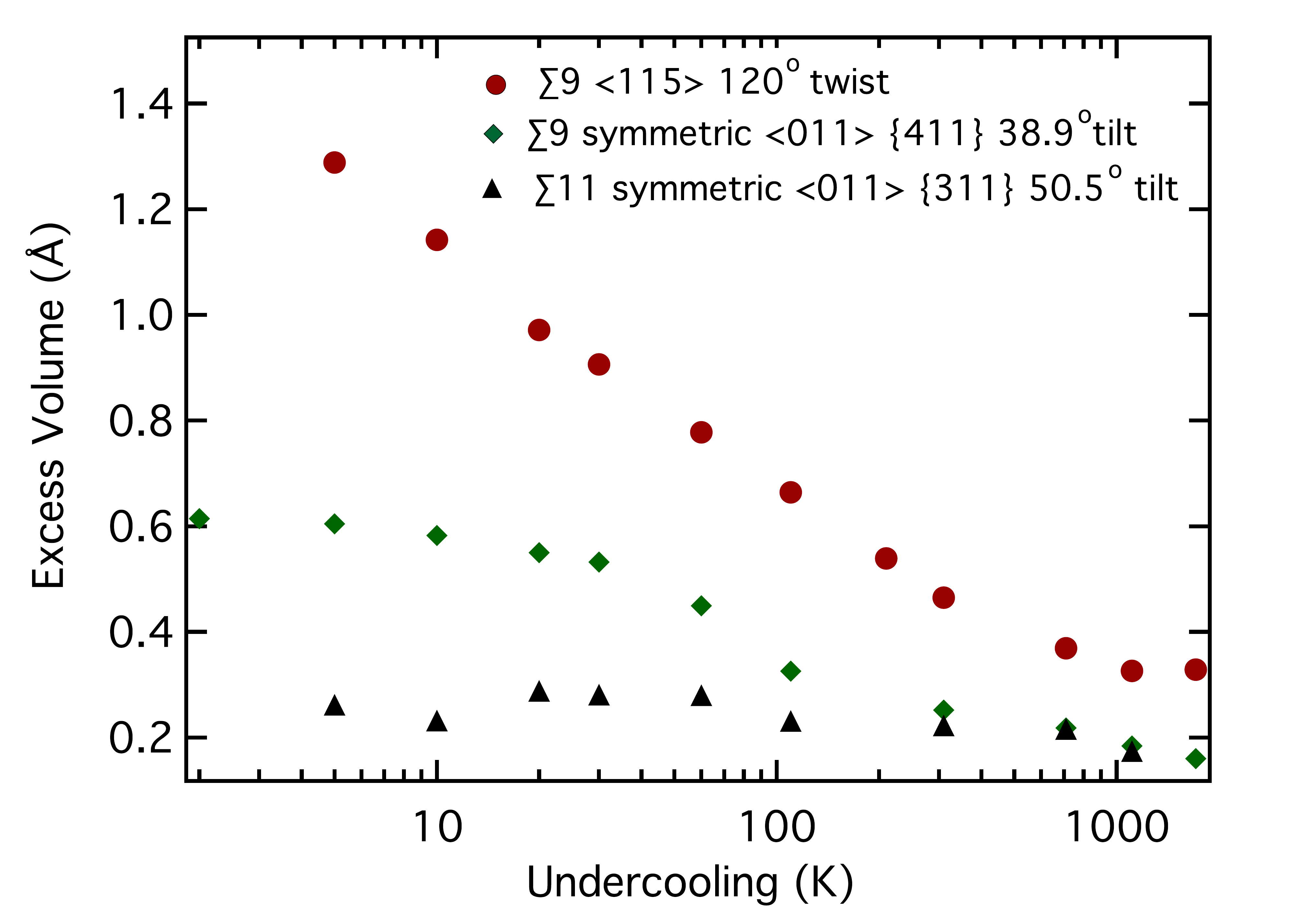}
\caption{\label{vol}Excess volume versus temperature plot for three 
different grain boundaries in Nickel. The red circles correspond to a 
$\Sigma$9 $\left\langle{1 1 5}\right\rangle$ twist grain boundary, 
the green diamonds correspond to a $\Sigma$9 symmetric $\left\langle{0 1 1}\right\rangle$$\left\{4 1 1\right\}$ tilt grain boundary
and the black triangles correspond to a 
$\Sigma$11 symmetric $\left\langle{0 1 1}\right\rangle$$\left\{3 1 1\right\}$ tilt grain boundary.
The three boundaries show very different behavior.}
\end{figure}

\begin{figure}
\includegraphics[scale=.35]{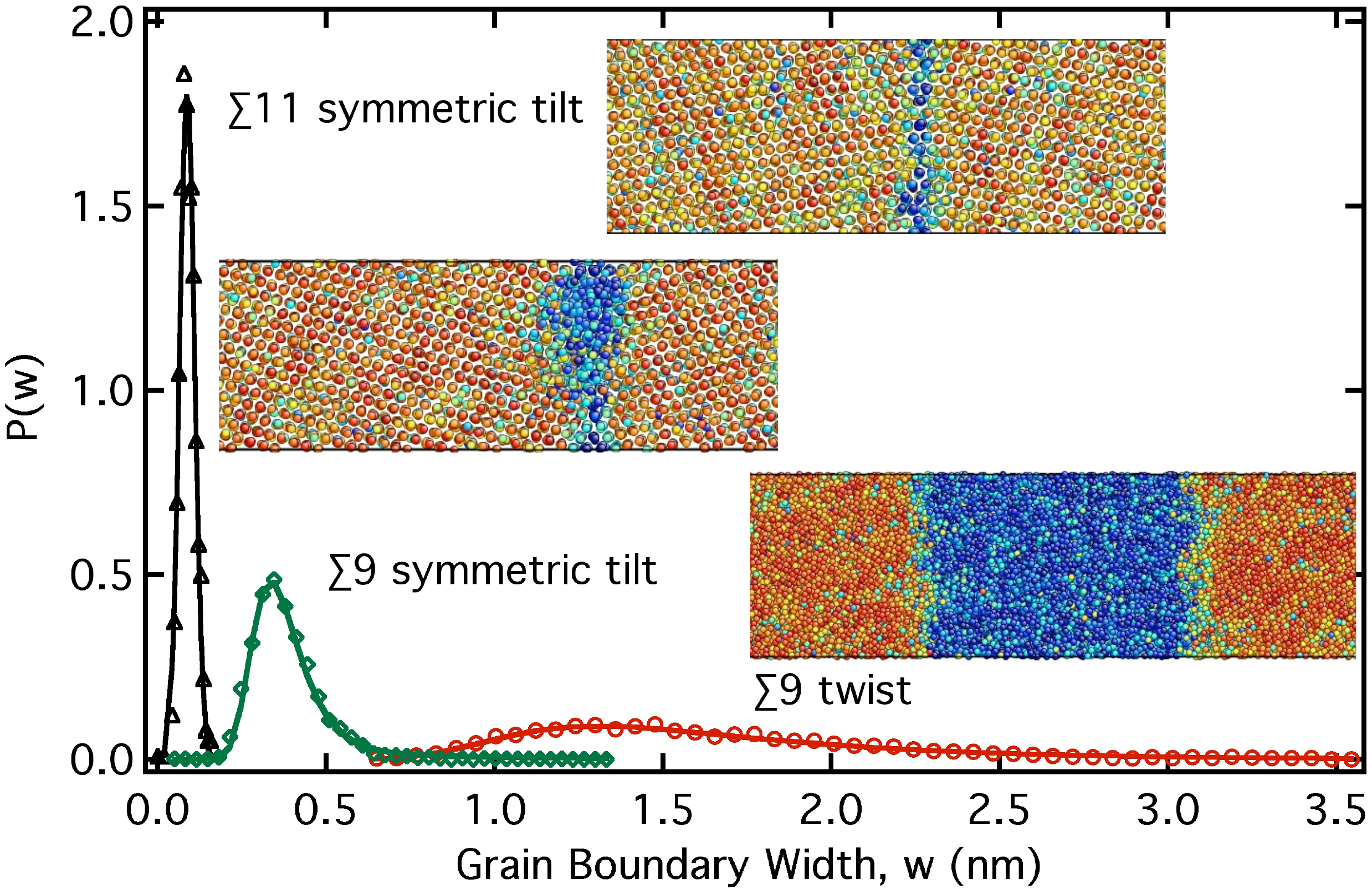}
\caption{\label{comp}The distribution function $P(w)$ vs $w$ 
from the MD simulations for the three different boundaries at 1708 K.   
The green diamonds and the red circles correspond to the $\Sigma$9 tilt and 
twist grain boundaries respectively. 
The$\Sigma$11 tilt boundary is represented by the black triangles. 
The lines represent a least square fit of the premelting models of 
Eqs.~\ref{eq} and \ref{eq4} for the $\Sigma$9 twist grain boundary,  
Eqs.~\ref{eq} and \ref{eq5} for the $\Sigma$9 tilt boundary and 
a gaussian fit for the $\Sigma$11 tilt boundary. 
The snapshots next to each figure correspond to the widest 
grain boundary for each boundary and are color coded according to the $\phi_i$ values. 
Lighter color (red in the on-line figure) indicates an atom with an environment corresponding to one of the  
two grains, darker color (blue in the on-line figure) represents atoms in some other environment.}                    
\end{figure}

The effect of temperature on the structure of the three boundaries 
considered in this work is illustrated in Figs.~\ref{vol} and 
\ref{comp}.  Figure~\ref{vol} plots the calculated excess volume 
versus the logarithm of the interface undercooling, and displays 
three qualitatively different types of behavior.  
The $\Sigma$9 twist boundary has an excess volume (represented by the red circles)
that continuously increases as the melting temperature is 
approached from below.  The behavior displayed by this boundary in 
Fig.~\ref{vol} is consistent with a logarithmic divergence of the 
interface width, as would be expected if the disjoining potential is 
of the form given by Eq.~\ref{eq4}.  At the opposite extreme, the 
$\Sigma$11 tilt boundary (represented by the black triangles) has an 
excess volume that is relatively small and depends only weakly on 
temperature.  The $\Sigma$9 tilt boundary (represented by the green 
diamonds) displays an intermediate behavior.  
The excess volume rises with increasing temperature at a rate that 
initially tracks that of the continously premelting $\Sigma$9 twist 
boundary.  At high temperature, the rate of increase of the excess 
volume decreases and the boundary maintains a finite excess volume 
at $T_M$.

Representative snapshots and calculated width histograms are shown 
in Fig.~\ref{comp} for each of the three boundaries at 1708 K (two 
degrees below $T_M$).  The width histogram corresponding to the 
$\Sigma$9 twist boundary is much broader than those for 
the other two boundaries.  This boundary samples 
relatively large widths, and $P(w)$ shows pronounced asymmetry with 
an extended tail to large $w$ values.  The $\Sigma$11 tilt boundary, 
by contrast, features a width histogram that is very narrow and is
centered on relatively small values of $w$.  The width distribution 
for this boundary is roughly Gaussian in shape, with no detectable 
asymmetry towards large values of $w$.  The width distribution for 
the $\Sigma$9 tilt boundary shows features intermediate between 
the other two.  The average width is larger than that of the 
$\Sigma$11 boundary, and the width distribution is considerably 
broader.  Compared to the $\Sigma$9 twist boundary, however, the 
asymmetry and the tail extending to larger widths is not nearly as 
pronounced in the width distribution for the $\Sigma$9 tilt 
boundary.

The MD snapshots in Fig.~\ref{comp} next to each histogram 
provide additional insights into the nature of the 
high-temperature structural disorder present in each of the three 
boundaries.  As in Fig.~\ref{prof}, the atoms in these snapshots 
have been colored based on the difference between values of the 
structural order parameters ($\phi_1$ and $\phi_2$) defined above.  
Lighter colors (red in the on-line figure) indicate atoms with an environment corresponding to one 
of the two grains, while darker color (blue in the on-line figure) denotes a disordered environment 
distinct from those in either of the grains.  The snapshots and 
width histograms show that the $\Sigma$9 twist boundary at 1708 K 
displays thick premelted layers, consistent with the continuous 
premelting behavior suggested by the excess volume.  The 
$\Sigma$11 tilt boundary is seen to display appreciable disorder 
only over a width that is on the scale of one atomic plane; this 
boundary is observed in the MD simulations to remain highly 
ordered up to $T_M$. The intermediate behavior of the $\Sigma$9 
tilt boundary is characterized by disorder that extends over a 
distance of several atomic planes.  An important feature of the 
disorder observed in this boundary is the nonuniform character
of the widths along the area of the boundary.  As illustrated 
in the snapshot of this boundary in Fig.~\ref{comp}, thick and extended 
regions of disorder up to a nanometer or more in thickness are often 
observed in part of the boundary, with much narrower, more ordered 
regions in between.  For the remainder of this section we will focus on 
the behavior of the two $\Sigma$9 boundaries, showing that the 
temperature dependence of the width histograms can be accurately 
described by the disjoining-potential formalism described above.

\begin{figure}[here]
\includegraphics[scale=.36]{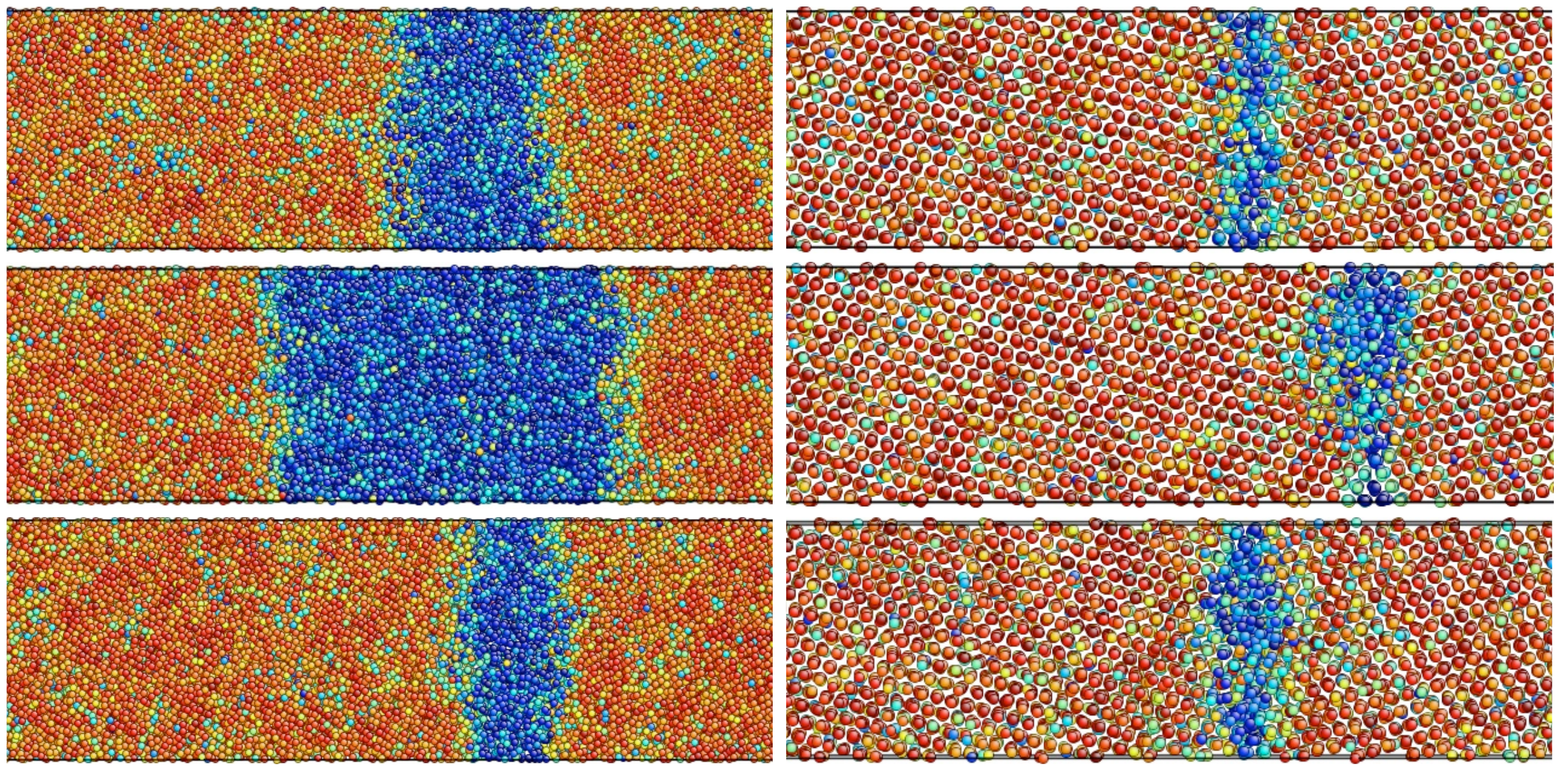}
\caption{\label{snapshots}Snapshots from a MD simulation at an undercooling 2 K illustrating 
the dynamic nature of the grain-boundary width.  The snapshots corresponding to the  
$\Sigma$9 tilt grain boundary are on the right of the panel. 
The snapshots to the left are from the $\Sigma9$ twist boundary discussed elsewhere \cite{hoyt09}. The atoms are colored by the difference between the two order parameters as described above.}
\end{figure}

Figure~\ref{snapshots} further illustrates the nature of the disorder 
present in the $\Sigma$9 twist and tilt boundaries.  The snapshots were 
obtained from a 40 ns MD simulation at 1708 K, and represent the largest,
smallest and an average value for the interface widths.  Results for
the $\Sigma$9 twist boundary are shown on the left and those for the
 $\Sigma$9 tilt boundary are on the right of Fig.~\ref{snapshots}.
The snapshots illustrate the highly dynamic nature of the boundary 
structures at 1708 K. The large fluctuations in interface width 
displayed by these boundaries forms the basis for the histogram 
analysis underlying calculations of the disjoining potential.

\begin{figure*}
\begin{center}
\includegraphics[scale=.7 ]{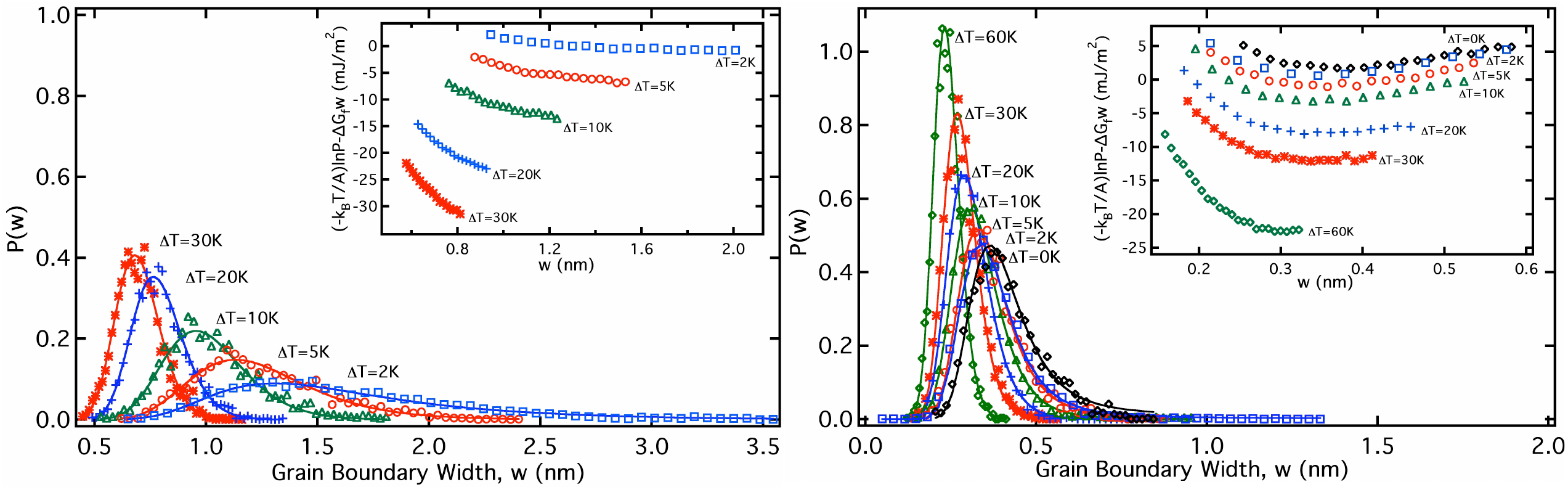}
\caption{\label{distr}The distribution function $P(w)$ vs $w$ from the MD 
simulations. The 
function 
corresponding to the  
$\Sigma$9 tilt grain boundary is in the right 
 of the 
panel. The 
 function 
to the 
left is from the $\Sigma$9 twist boundary discussed elsewhere \cite{hoyt09}. 
The lines represent least square fits of the premelting model of 
Eqs.~\ref{eq} and \ref{eq4} for the $\Sigma$9 twist grain boundary and 
Eqs.~\ref{eq} and \ref{eq5} for the $\Sigma$9 tilt boundary.}
\end{center}
\end{figure*}

The calculated width histograms for both $\Sigma$9 boundaries are
displayed over a range of interface undercoolings in Fig.~\ref{distr}. 
The histograms on the left and right in Fig.~\ref{distr} correspond to 
the $\Sigma$9 twist and tilt boundaries, respectively.  For both of the 
boundaries the histograms become broader as the bulk melting temperature 
is approached (i.e., with decreasing undercooling).  For the $\Sigma$9 
twist boundary the system is observed to melt completely over the 
course of the 20 ns runs at the bulk melting temperature, and 
histograms can be obtained for this boundary only for finite values of 
the undercooling.  By contrast, the width of the $\Sigma$9 tilt 
boundary remains finite at $T_M$ and the width histogram can be 
calculated at zero undercooling for this boundary, as shown in 
Fig.~\ref{distr}.  The inset in each of the panels shows the quantity
$-k_B T_i \log{[P(w,T_i)]} / A - \Delta G_f w$ versus $w$ for all of
the interface temperatures.  The data for the $\Sigma$9 tilt boundary
shows clear evidence of an attractive minimum in the disjoining 
potential, as will be discussed further below.

The solid lines in Fig.~\ref{distr} represent the least square fits of 
the MD data for $P(w)$ to the disjoining-potential formulas given by 
Eqs.~\ref{eq1}, \ref{eq} and \ref{eq4} for the $\Sigma$9 twist boundary, 
and Eqs.~\ref{eq1}, \ref{eq} and \ref{eq5} for the $\Sigma$9 tilt boundary.  
For the $\Sigma$9 twist boundary the potential parameters obtained from 
the separate fits to the data for the individual temperatures span the 
range $\delta$=0.25 to 0.29 nm, and 
$\Delta \gamma$=101 to 150 mJ/m$^2$.  For the $\Sigma$9 tilt boundary 
the fitted values for $\delta_1$ and $\delta_2$ ranged between 0.142 to 
0.144 and 0.143 to 0.144 nm, respectively, $\Delta_1 - \Delta_2$ spanned 
the values 103 to 163 mJ/m$^2$, and $\Delta_2 / \Delta_1$ took values 
between 1.003 and 1.006.  With these fitted potential parameters, the 
disjoining potential for the $\Sigma$9 tilt boundary exhibits a weak 
minimum at a width $w_m$ with values ranging between 0.32 and 0.36 nm, 
and a depth relative to $2 \gamma_{SL}$ varying between -7 and -11 
mJ/m$^2$. The fact that the MD calculated width histograms can be 
accurately described by disjoining potentials with a relatively narrow 
range of fitted potential parameter values indicates that the formalism 
described in the previous section represents a valid model for 
describing the temperature dependence of the structural disorder observed 
in these boundaries.

\begin{figure}[here]
\includegraphics[scale=.35]{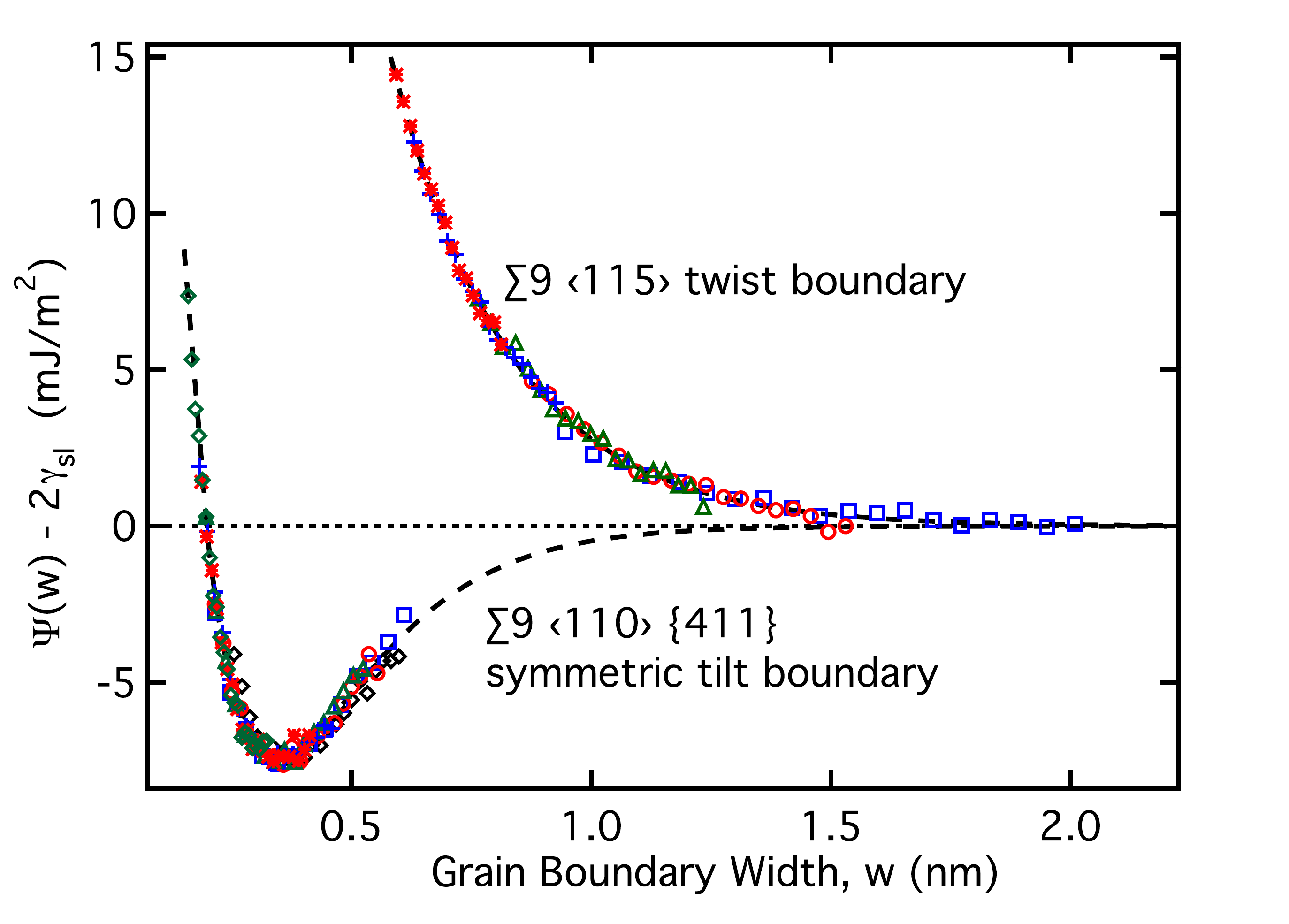}
\caption{\label{disjoin}An illustration of the histogram method 
used to extract the disjoining potential. It shows the merged data 
from individual histogram data used to reproduce the complete 
disjoining potential $\Psi(w)$ for the $\Sigma$9 twist and the 
$\Sigma$9 tilt boundaries.  The solid line is the best fit to the 
exponential decay given in Eq.~\ref{eq4} for the  $\Sigma$9 twist boundary, 
and a double exponential function in Eq.~\ref{eq5} for the  $\Sigma$9 tilt grain boundary. }
\end{figure}

In order to refine the calculation of the disjoining potential for
the  $\Sigma$9 twist and tilt boundaries, we employ the histogram
analysis described in the previous section, involving a refinement
of both the shift parameters $a_i$ and the potential parameters in
Eqs.~\ref{eq4} and \ref{eq5} for the twist and tilt boundaries,
respectively.  The resulting fits are shown in Fig.~\ref{disjoin}
and correspond to the following values for the potential parameters: 
$\delta=0.25$ nm and $\Delta\gamma=156$ mJ/m$^2$ for the $\Sigma$9
twist boundary, and
$\Delta_1 - \Delta_2 = 103$ mJ/m$^2$, $\Delta_2 / \Delta_1 = 1.003$,
$\delta_1=0.1471$ nm, $\delta_2=0.1474$ nm for the 
$\Sigma$9 tilt boundary.  These parameter values are consistent with
the values given above from the independent fits.  The excellent 
agreement of the fits with the MD data in Fig.~\ref{disjoin} again
indicates that the disjoining-potential formalism represents an 
accurate framework for modeling the premelting behavior of these
$\Sigma$9 boundaries.  The analysis used to obtain the results in 
fits in Fig.~\ref{distr} assumed a melting point of $T_M=1710$ K.  If 
the melting temperature is changed even by one degree,i.e., 
$T_M=1709 K$ then poor fits to $P(w)$ are obtained for the data at the 
lowest undercoolings.

The calculated disjoining potentials in Fig.~\ref{disjoin} are 
characterized by the following features.  For the $\Sigma$9 tilt
boundary the disjoining potential has a minimum at a finite width, 
$w_m$ ,which corresponds to the average equilibrium grain boundary 
width at the melting temperature. The potential is repulsive for 
$w<w_m$ and attractive for $w>w_m$.  In contrast, the $\Sigma$9 
twist boundary features a purely repulsive disjoining potential that
is well modeled by exponential form given in Eq.~\ref{eq4}, as 
expected based on the logarithmic divergence of the excess volume 
calculated for this boundary (see Fig.~\ref{vol}).

\section{\label{con}Summary and Discussion}
In this paper we have presented a detailed description of 
a method for computing the disjoining potential for grain-boundary
premelting and grain coalescence from an analysis of width 
fluctuations measured in equilibrium MD simulations.  The 
approach has been applied to two grain boundaries in an EAM model
of elemental Ni.  For the $\Sigma$9 twist boundary, which features 
an excess volume that increases logarithmically as $T_M$ is 
approached from below, the measured width histograms are consistent 
with a disjoining potential that decays exponentially, with a decay 
length of approximately $0.25$ nm and a maximum value at zero width 
of approximately 160 mJ/m$^2$ relative to $2 \gamma_{SL}$.  For the 
$\Sigma 9$ tilt boundary, which features an excess volume that 
remains finite at $T_M$, the measured width histograms are shown 
to be consistent with a disjoining potential that features a weak 
attractive minimum at $w_m \approx 0.34$ nm, with a depth relative 
to $2 \gamma_{SL}$ of approximately -8 mJ/m$^2$.

It is interesting to compare the results of the present work with 
previous studies of grain-boundary premelting based on MD 
simulations, phase-field theory and PFC calculations.  The 
exponential form of the disjoining potential for the $\Sigma$9 twist 
boundary presented here and in Ref.~\cite{hoyt09} is qualitatively 
consistent with the results of previous investigations where a  
diverging grain-boundary width has been observed for high-energy 
boundaries in a variety of different systems (e.g., 
\cite{brough86, brough98,kaski,kikuchi}).
The disjoining potential calculated here for the $\Sigma$9 tilt 
boundary features a weak attractive minimum at $w_m$, and is 
replusive and attractive for smaller and larger values of $w$, 
respectively.  This form for the disjoining potential corresponds 
to a grain-boundary whose width increases with $T$ at low 
temperature, but remains finite at and above $T_M$ (up to some 
maximum superheating).  This type of behavior for the temperature 
dependence of the grain boundary width is qualitatively similar to 
that found in recent MD studies for one of three twist boundaries 
in Si \cite{kaski} and a symmetric tilt boundary in Cu 
\cite{williams}.  

As discussed in the introduction section disjoining potentials with 
attractive minima, qualitatively similar to that calculated here
for the $\Sigma$9 tilt boundary, were obtained in PFC calculations
by Mellenthin et al.\cite{mellenthin08} for low-angle tilt boundaries
in a model two-dimensional hexagonal system.  The boundaries which
displayed this type of disjoining potential had highly non-uniform 
interface structures, containing premelted regions localized on 
grain-boundary dislocation cores and separated by well ordered 
regions which we will refer to as solid ``bridges".  For these
structures, the width corresponding to the minimum in the 
disjoining potential, which represents an average over the area of 
the boundary, corresponds to the thickness of an equivalent 
uniform layer with the same amount of ``liquid" as contained in 
the premelted cores.  This area-averaged width can be quite 
small (e.g., on the order of the atomic spacing) even though the 
radius of the premelted cores is much larger.

\begin{figure*}
\begin{center}
\includegraphics[scale=0.6]{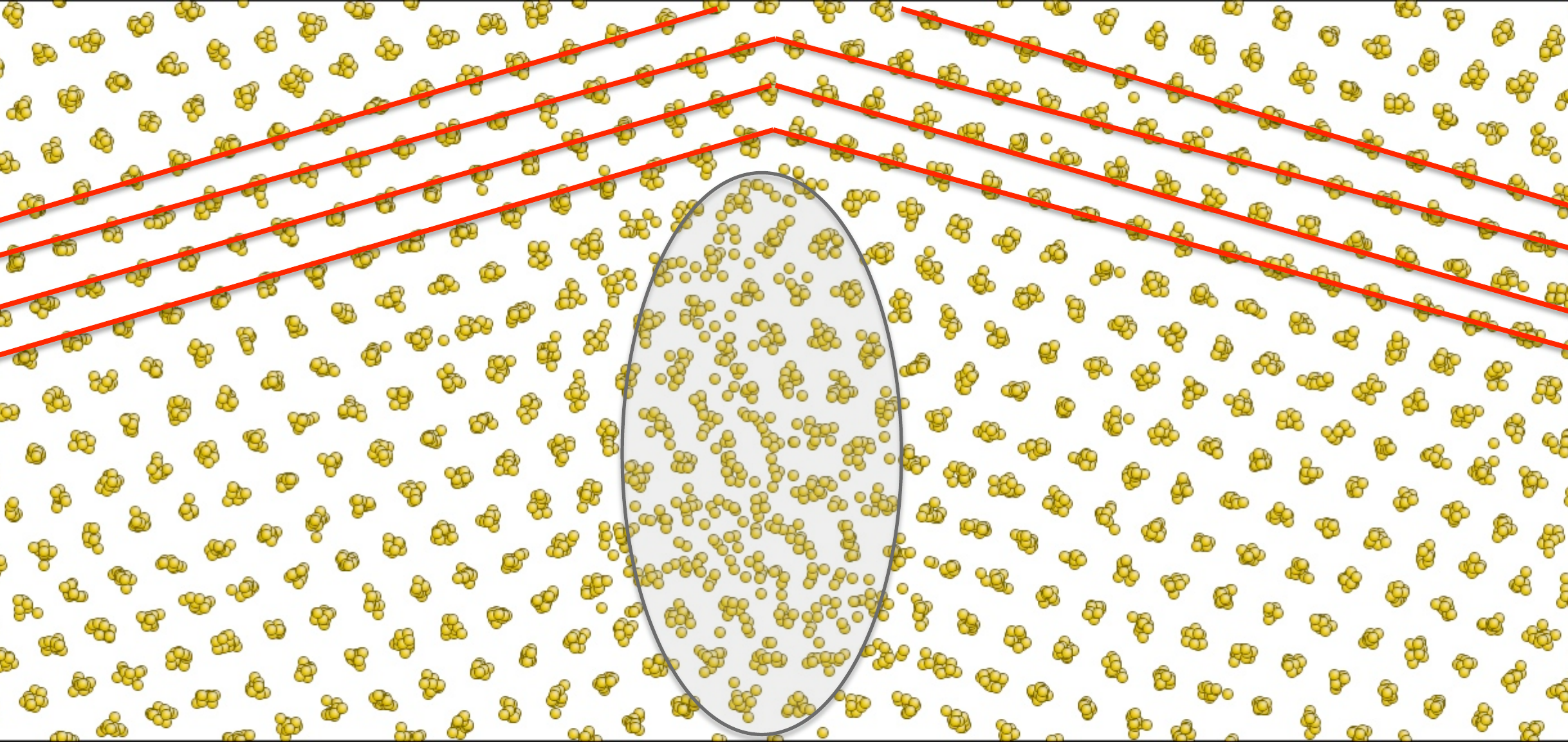}
\caption{\label{bridge}  Snapshot of the $\Sigma$9 tilt boundary at 1708 showing 
the non-uniform structural disorder in the grain boundary.  The grey ellipse shows 
the regions with high disorder while the red lines highlight the ordered bridges.}
\end{center}
\end{figure*}

It is interesting to compare the structures observed in the PFC 
calculations of Mellenthin et al. \cite{mellenthin08} with those 
for the $\Sigma$9 tilt boundary considered in the present study.  
Visual inspection of the numerous snapshots showed that the disorder 
in this boundary is indeed often highly non-uniform, as illustrated 
in Fig.~\ref{bridge}.  

This figure shows a representative snapshot 
from a simulation at $T=1708$ K, viewed down the tilt axis.  In 
Fig.~\ref{bridge}, the structural disorder in the boundary is 
clearly nonuniform - the region of high structural disorder is 
highlighted by the grey ellipse, while the red lines connected across 
the boundary plane highlight the ordered solid bridges.  Thus, even 
though the $\Sigma$9 boundary studied in the present work has a 
misorientation angle (38.9 degrees) which is far too large for its 
structure to be described in terms of separated grain-boundary 
dislocation cores, the non-uniform nature of the structural disorder
and the presence of solid ``bridges" is qualitatively very similar
to the types of structures observed in Ref.~\cite{mellenthin08}.

It should be emphasized, however, that the structure of the
$\Sigma$9 boundary observed in the MD simulations is highly dynamic
(c.f., Fig.~\ref{snapshots}) such that the solid bridges form and 
disappear rapidly on the time scale of the simulations.  This 
behavior could have interesting consequences for the shear response
of such boundaries, as the solid bridges are expected to offer 
enhanced resistance to shear which would otherwise be expected
to be very limited for a premelted grain boundary 
(e.g., \cite{brough98}).  The shear response of such 
boundaries would thus be an interesting topic for future MD studies.

\section{Acknowledgments}
Work at Northeastern, UC Davis and Berkeley was supported by the Director,
Office of Science, Office of Basic Energy Sciences, Materials Sciences and
Engineering Division, of the U. S. Department of Energy (DOE), under
contracts DE-FG02-07ER46400, DE-FG02-06ER46282, and
DE-AC02-05CH11231.  JJH acknowledges financial support from a Natural
Sciences and Engineering Research Council (NSERC) of Canada Discovery grant.
All the authors acknowledge support from the DOE Computational Materials Science
Network program.  This research used resources of the National Energy Research
Scientific Computing Center, which is supported by the Office of Science of the U. S.
Department of Energy under Contract No. DE-AC02-05CH11231.  We also acknowledge helpful discussions with R. G. Hoagland.

\end{document}